\documentclass[11pt]{article}
\usepackage{authblk}

\usepackage{amsmath,graphicx}
\usepackage[round,numbers, sort&compress]{natbib}
\usepackage{todonotes}

\makeatletter
\def\tagform@#1{\maketag@@@{[#1]\@@italiccorr}}
\makeatother
\usepackage{setspace}
\usepackage[margin=1in]{geometry}
\usepackage{amssymb}
\usepackage{epstopdf}
\usepackage{subfigure}
\usepackage{color}
\usepackage[normalem]{ulem}
\usepackage{amsthm}
\usepackage{algorithm}
\usepackage{algorithmic}
\usepackage{xr}
\usepackage{hyperref}

\title{\uppercase{ \bf \large \center A General Algorithm for Compensation of Trajectory Errors: Application to Radial Imaging}}

\author[1]{\it \small Merry Mani}
\author[1]{\it \small Vincent Magnotta}
\author[2]{\it \small Mathews Jacob}
\affil[1]{\it \small Department of Radiology, University of Iowa, Iowa City, Iowa}
\affil[2]{\it \small Department of Electrical and Computer Engineering, University of Iowa, Iowa City, Iowa}

\begin{document}
\maketitle

\vspace{30mm}
\noindent
Correspondence to :\\
Merry Mani\\
L420 Pappajohn Biomedical Discovery Building\\
169 Newton Road\\
Iowa City, Iowa, 52242\\
email: merry-mani@uiowa.edu \\
phone number: (319) 335-5183.\\
\\
Word count : about 2800\\
Figures+ tables count : 5\\
\\
Running title: Annihilating filter based k-space shift correction \\

\newpage
\noindent{ \bf Abstract}

Purpose: To reconstruct artifact-free images from measured k-space data, when the actual k-space trajectory deviates from the nominal trajectory due to gradient imperfections. 

Methods: Trajectory errors arising from eddy currents and gradient delays introduce phase inconsistencies in several fast scanning MR pulse sequences, resulting in image artifacts. The proposed algorithm provides a novel framework to compensate for this phase distortion. The algorithm relies on the construction of a multi-block Hankel matrix, where each block is constructed from k-space segments with the same phase distortion. In the presence of spatially smooth phase distortions between the segments, the complete block-Hankel matrix is known to be highly low-rank. Since each k-space segment is only acquiring part of the k-space data, the reconstruction of the phase compensated image from their partially parallel measurements is posed as a structured low-rank matrix optimization problem, assuming the coil sensitivities to be known. 

Results: The proposed formulation is tested on radial acquisitions in several settings including partial Fourier and golden-angle acquisitions. The experiments demonstrate the ability of the algorithm to successfully remove the artifacts arising from the trajectory errors, without the need for trajectory or phase calibration. The quality of the reconstruction was comparable to corrections achieved using the Trajectory Auto-Corrected Image Reconstruction (TrACR) for radial acquisitions. 

Conclusion: The proposed method provides a general framework for the recovery of artifact-free images from radial trajectories without the need for trajectory calibration.

\vspace{2em}

\newpage

\noindent{\bf INTRODUCTION}

\noindent Several different k-space sampling strategies, each with unique benefits, are used in MRI.  Echo-planar imaging (EPI) trajectories that scans the k-space in a single sweep, is routinely used in high temporal sampling applications such as functional MRI and diffusion MRI\cite{Poustchi-Amin2013, Poser2007,Jaermann2004}. Likewise, radial schemes, that sample the k-space center during each projection is ideally suited for accelerated imaging  \cite{Peters2006}. Center-out radial methods with their short read-outs has enabled zero echo-time applications such as lung and sodium imaging \cite{Nagel2009,Johnson2013}. Despite their benefits, a key problem that affects the performance of the above fast-scan methods is gradient fluctuations. Several hardware imperfections exist in the MR system that limit the exact realization of a prescribed gradient waveform on the MR hardware. Among these, eddy currents and gradient timing delays are understood as the leading source of gradient fluctuations \cite{Boesch1991,Aldefeld1998}. The gradient fluctuations result in the actual k-space encoding to be different from the expected trajectory leading to reconstruction errors. The manifestation of the artifacts in the images will vary based upon the trajectory \cite{Smith2010}.

\par The classical approach to compensate for the trajectory errors is to measure the deviated k-space trajectory either using specialized hardwares \cite{Ianni2016} or employing calibration scans \cite{Xu2010}. The calibration-based methods have been proposed for radial \cite{Block} and EPI trajectory correction \cite{Bruder1992}. However, these methods are not effective in dealing with temporal variations of the trajectory arising from gradient heating. Subject motion between the calibration scans and the imaging experiment can also lead to poor corrections. Self-calibrating trajectory correction methods are immune to these problems. 

\par Self-calibrating methods based on iterative phase cycling were originally introduced for EPI \cite{Clare1990,Foxall1999,Chen2011} and later extended to radial acquisitions \cite{Deshmane2016,Wech2015}. The EPI phase cycling methods reconstruct a series of images using different choices of phase errors between the odd and even lines and the phase which minimized some criteria was used in the final reconstruction. Likewise, Deshmane et al. \cite{Deshmane2016}  iteratively shift the radial data until it maximizes the root sum-of-squares (SOS) DC signal of each individual projection, while Wech et al. \cite{Wech2015} iteratively shift the radial trajectory in a direction that provides the best data fit. Recently, these methods were generalized by an optimization scheme for multiple trajectories; this approach relies on a basis expansion for eddy current and gradient non-linearities, along with a gradient descent optimization of the trajectory parameters \cite{Ianni2016}. A challenge with the above approach is the non-quadratic nature of the optimization scheme, associated with the need for careful initialization of the parameters to ensure convergence.

\par Recently, we proposed a self-calibrating Nyquist ghost correction strategy for EPI reconstruction \cite{Mani2016a,Mani2016c}. Here, we recovered the missing k-space samples in the odd and the even blocks after the odd and the even lines of the EPI k-space were separated from each other. The images corresponding to the odd/even datasets  differ only by a phase error, induced by the k-space shifts. We observed that these phase relations manifest as annihilation conditions which enabled the formulation of a block-Hankel matrix, created from these datasets, that is heavily low-rank. We rely on Hankel structured low-rank matrix completion (SLRMC) to fill in the missing entries in the datasets. This phase-compensating reconstruction is termed as MUSSELS\cite{Mani2016b,Mani2016a}. In the current work, we generalize the above idea to trajectories beyond EPI. We show the applicability of the method for correction of trajectory errors in multiple radial-based acquisitions where we do not know the trajectory shifts a priori.  A preliminary account of this work was presented at the 2017 ISMRM \cite{ManiMerry2017}. \\

\noindent{ \bf METHODS}\\
\noindent{  \bf Modeling of trajectory errors as phase modulations }\\

\par Gradient fluctuations arising from hardware imperfections result in trajectory shifts in several acquisitions. The proposed method is based on the idea that, in many cases, the effect of k-space shifts can be modeled as phase modulations in the data.  Let us denote the ideal image that is not affected by trajectory errors and its k-space data by $\rho(\bf r)$ and $ \widehat \rho~[\bf k]$ respectively for an arbitrary trajectory; $\bf r$ and $\bf k$ being the spatial and k-space position index vectors.  In the presence of trajectory errors, the k-space lines of $ \widehat \rho~[\bf k]$ will be shifted by various amounts. We propose that the shifted k-space data can be segmented into $N$ segments, where the data in each segment has the same shift. 
\begin{equation}
 \widehat{I_i}[{\bf k}] =\widehat \rho~[{\bf k}+\Delta {\bf k}_i], ~~~ i=1:N.
 \end{equation}
 Here, $\Delta {\bf k}_i$; $i=1:N$ denote the discrete shifts of each segment.  An illustration with (N=) 4 different shifts affecting the k-space lines of $ \widehat \rho~[\bf k]$ is provided in Figure 1a and our proposed splitting is illustrated in Figure 1b. Note that each of the segments are not fully sampled in k-space. If they were fully sampled, we can recover the images $I_{i}(\mathbf r); i=1,..N$ by simply taking the inverse Fourier transform of the full k-space data as:
\begin{equation}
I_i({\bf r})={\cal F}^{-1}\{~\widehat{I_i}[{\bf k}] ~ \}={\cal F}^{-1}\{ ~\widehat \rho~[{\bf k}+\Delta {\bf k}_i]  ~\}=\rho({\bf r})e^{j2\pi\Delta k_i}, \forall i=1:N.
 \end{equation}
Note that the images $I_i({\bf r})$ are phase-modulated versions of the underlying image $\rho(\bf r)$. Since each of the datasets $\widehat{I_i}[{\bf k}]$ are not fully sampled, we denote the measurement process as 
 \begin{equation}
\mathcal A_i(I_i) = \widehat {\mathbf y}_i; ~~ i=1,..,N.
\end{equation}
Here, $\mathcal A_i$ includes the coil sensitivity weighting, Fourier transform, as well as the sampling operator that selects the acquired samples for the $i^{\rm th}$ segment. Similarly, $\widehat {\mathbf y}_i$ denotes the acquired measurements of the dataset corresponding to the $i^{\rm th}$ segment. We propose to recover the images $I_i$ using the phase relations between them.\\

 \noindent{\bf Trajectory Corrected Reconstruction based on Structured Low-Rank Matrix Completion}\vspace{0.5em}

 \par In \cite{Mani2016b}, we showed that there exists annihilation relations $~\widehat{I_i} * h_i - \widehat{I_j} * h_j = 0$ between the Fourier samples of the phase-modulated images $ I_i; i=1,..,N$. Here, * denotes  convolution and $h_i$ and $h_j$ are 2-D finite impulse response filters. These annihilation conditions can be compactly expressed as null-space conditions on the below block-Hankel matrix: 
 \begin{equation}
\label{Hankel_radial}
{\bf{{H}}(\hat{I})}=\begin{bmatrix} {\bf{\cal H}}(\widehat{{{I}}_1}) & {\bf{\cal H}}(\widehat{{{I}}_2}) & ... & {\bf{\cal H}}(\widehat{{{I}}_{N}}) \end{bmatrix}.
\end{equation}
Here, each block $~\mathcal H(\widehat{{I}}_i)$ is a Hankel matrix, formed out of the fully sampled k-space data of the phase-modulated images \cite{Shin2014}. ${\bf{H}}$ is the operator that converts the k-space data of the phase-modulated images $\mathbf I = [I_1,..,I_N]$ into a multi-block Hankel matrix (please see supporting information for details). The null-space conditions induced by the annihilation relations imply that ${\bf{{H}}(\hat{I})}$ is a low-rank matrix. 

\par Since each of the phase-modulated images $I_i$ are not fully sampled, several entries of the matrix $\mathbf H$ are unknown. These entries can be recovered using the SLRMC approach as in \cite{Shin2014,Haldar2015,Jin2017,Lee2016a}. However, in contrast to the parallel imaging setting in \cite{Shin2014,Haldar2015,Jin2017,Lee2016a}, there are no shared samples of k-space between $\widehat {I_i}[{\bf k}]; i=1,..,N$ (Figure 1b). Hence, the direct extension of the approaches in \cite{Shin2014,Haldar2015,Jin2017,Lee2016a} to our setting will be ill-posed. Specifically, one could fill the missing entries in each matrix with the samples from other ones. This will also result in a heavily low-rank matrix since the k-space data for all blocks are the same \cite{Lobos2017}. To avoid this pitfall, the MUSSELS algorithm is formulated as: \begin{equation}
 \label{Recon}
\{\tilde { I_j},j=1,..,N\}=  \text{argmin}_{\{ { I_j},j=1,..,N\}} \left(\sum_{j=1}^{N}~\|\underbrace{\mathcal{A}_j( I_j)-\mathbf {\hat{y}}_j\|^2_{\ell_2}}_{\text{\footnotesize{data consistency}} }+ ~~~\lambda\underbrace{||{\bf{{H}}(\hat{I})}||_*}_{\text{\footnotesize{low-rank property}}  }\right), \vspace{-2mm}
\end{equation}
which make use of an additional data-consistency constraint \cite{Mani2016b,Mani2016a}.  Specifically, filling the missing entries with the data from others will violate this data-consistency. 

\par The reconstruction given in Eq. \ref{Recon} with the forward model operator $\mathcal{A}$ appropriately defined for the specific trajectory gives the general trajectory correction method based on the idea of phase compensation.\\

 \noindent{\bf Radial trajectories }\vspace{0.5em}
 
\par The key step in achieving effective trajectory correction based on the above method is grouping the k-space lines that experience similar errors. While this step is straightforward for some trajectories (e.g. EPI), a well-defined separation may not be easily available for several trajectories, as in the case of conventional or golden angle radial acquisitions. The challenge is in identifying this grouping. For both conventional and golden angle radial  trajectories, it has been shown that the trajectory shifts vary as a function of the angle of the spoke \cite{Peters2003,Ianni2016}. Therefore, we propose a grouping that is based on the angle of the radial spokes for these trajectories. Specifically, the radial spokes with similar angles can be assumed to have similar trajectory errors. This enables us to generalize the phase-compensated reconstruction based on SLRMC for correction of trajectory errors in radial-based acquisitions also.  

\par Denoting the measured radial k-space data on the non-Cartesain grid by $\bf \hat{ y}$, we split the radial spokes of a given frame into $N$ segments $\mathbf {\hat {y}}_i,  i=1:N$ based on their angles. To keep consistent notations for Cartesian and non-Cartesian trajectories, we denote the images and the k-space data on the Cartesian lattice corresponding to the N segments of the radial trajectories as $ I_i$ and $\widehat  I_i~; i=1,..,N$, respectively. 
The operator, $\mathcal A$, for the case of radial trajectories, is given by  ${{\cal{F}}_{j}}^{nu}~\circ~\bf{{\cal{S}}}$ where  $\bf{{\cal{S}}}$ represent coil sensitivities and ${{\cal{F}}_j}^{nu}$ represent the non-uniform fast Fourier transform (NUFFT) to the grid corresponding to $\mathbf {\hat {y}}_j$. $\mathcal H(\widehat{{{I}}_j})$ is computed using the Fourier samples of $ I_j$ on the Cartesian grid.  

\par The proposed reconstruction in Eq. \ref{Recon} was implemented in MATLAB 2016a (The Mathworks, Natick, MA) on a desktop PC with an Intel i7-4770, 3.4 GHz CPU with 8 GB RAM. The different steps involved in the proposed reconstruction are illustrated in Figure 2, which is solved using an augmented Lagrangian scheme (derived in the supporting information). The NUFFT is implemented as in \cite{Jacob2009}. The conjugate gradient updates use 20 iterations or a tolerance of 1e-15, with no pre-conditioning.
\\

\noindent{ \bf { Validations }}\vspace{0.5em}

\par	{\it Simulation study}: We first demonstrate the trajectory correction employing MUSSELS (referred to as T-MUSSELS) using simulations. We used a 32-channel brain data \underline{(\href{https://people.eecs.berkeley.edu/~mlustig/Software.html}{available as part of the ESPIRiT toolbox})} for this experiment. We simulated trajectory errors into a nominal radial trajectory by explicitly shifting the spokes of the trajectory. The shifts were generated as a function of the spoke angle. The k-space data to be used for testing was synthesized by performing a NUFFT of the above brain data using the corrupted trajectory. Using this simulated experiment, we also demonstrate that it is possible to recover the corrected image for a given trajectory error using more than one scheme of segmentation for the case of radial trajectories.  
 
\par {\it Experimental datasets}: Trajectory correction using T-MUSSELS was also validated using phantoms and in-vivo data from cardiac and neuroimaging studies. To demonstrate the generalizability of the above method, the experimental data was collected using a variety of different radial acquisitions. The cardiac data was acquired using a short-axis $512 \times 253 $ radial scan on a Siemens 3T scanner with 3-channel acquisition. The golden angle (GA) radial brain data is distributed as part of the TrACR package \cite{Ianni2016}. This 14-channel data consist of 201 radial spokes with 256 points per spoke. The phantom data was collected with a partial Fourier radial acquisition on a GE 3T scanner. This 32-channel data consist of 256 radial spokes with 144 points per spoke. 

\par Coil sensitivity maps are required for all the reconstructions proposed here. The coil sensitivity maps were estimated from the data itself in all experiments, by taking the data from a small region around the center of k-space using the ESPIRiT method \cite{Uecker2014}.\\

\noindent{ \bf RESULTS}\vspace{0.5em}

\par{ \it Simulations}: The results of the simulated experiment are provided in Figure \ref{fig:sim_epi_radial}. The shifted radial trajectory is shown in Figure \ref{fig:sim_epi_radial}a.  A CG-NUFFT reconstruction \cite{Jacob2009} of the shifted k-space data with the nominal radial trajectory show severe artifacts (Figure \ref{fig:sim_epi_radial}c). We used T-MUSSELS to correct for the trajectory errors using the two schemes shown in Figure \ref{fig:sim_epi_radial}b: by splitting the radial data into (i) 8 segments, and (ii) 6-segments.  The resulting reconstructions are shown in Figure \ref{fig:sim_epi_radial}c.  As noted from the results, it is possible to achieve good reconstruction with more than one splitting scheme for the case of the radial trajectory using the proposed method. 
  
 \par { \it In-vivo results}: Figure \ref{fig:cardiac_radial} shows the reconstruction from two frames of a cardiac data that was acquired using a radial trajectory. Figure \ref{fig:cardiac_radial}a show respectively the nominal radial trajectory and the trajectory that was split into 6 segments to be used with the proposed reconstruction. The left column in Figure \ref{fig:cardiac_radial}b-c shows the CG-NUFFT reconstruction using the nominal trajectory from two cardiac frames. A 4x4 filter was used for the 6-block Hankel matrix construction. The trajectory corrected reconstructions using T-MUSSELS from this three-channel acquisition show reduction in streaking and improved reconstruction in many regions. For comparison, the correction achieved using the TrACR method is also provided. Notably, the T-MUSSELS reconstruction provide improved contrast in the heart region (boxed region). 
 
 \par Figure \ref{fig:GA_radial}a shows the reconstruction of a GA radial data. Since the consecutive spokes of a GA radial are not close in angle, we reordered the radial data based on the angle of the spokes so that the data obtained from spokes with similar angles are grouped together. The new grouping, which segmented the spokes into 8 segments are shown in figure \ref{fig:GA_radial}a(ii). Figure \ref{fig:GA_radial}a(iii) shows the CG-NUFFT reconstruction that assumes the nominal trajectory shown in Figure \ref{fig:GA_radial}a(i). Figure \ref{fig:GA_radial}a(iv)-(v) show the reconstructions using the TrACR and T-MUSSELS respectively, which show improved reconstructions. 

\par Figure \ref{fig:GA_radial}b shows T-MUSSELS reconstruction validated on a phantom data that was acquired using a partial Fourier radial trajectory. Here, the spokes were split into 4 segments based on their angles as shown in Figure \ref{fig:GA_radial}b(ii). Figure \ref{fig:GA_radial}b(iii) shows the CG-NUFFT reconstruction that assumes the nominal trajectory given in Figure \ref{fig:GA_radial}b(i). The effect of trajectory errors are visible in this reconstruction. Figure \ref{fig:GA_radial}b(iv) shows the trajectory corrected reconstructions using the TrACR method. Even though there is significant reduction in artifacts, residual errors can still be seen in the resulting reconstruction. Figure \ref{fig:GA_radial}b(v) shows the proposed reconstruction using the nominal trajectory which shows improved results compared to the TrACR method. To quantify the difference, we computed the background-to-object signal intensity ratio similar to the ghost-to-signal ratio (GSR) that is typically computed for EPI based reconstructions \cite{Lee2016a}. As observed, the GSR is much lower in T-MUSSELS compared to TrACR.\\

\noindent{ \bf Discussion}\vspace{0.5em}

\par The proposed method provides a general self-calibrating two-dimensional phase compensation framework for the correction of trajectory errors. This work adds to the recent literature on structured low-rank based calibration-free strategies, which are emerging as powerful tools in a variety of applications including parallel MRI \cite{Morrison2007,Shin2014,Haldar2014,Haldar2015,Jin2017, Ye2015,Lee2016a} and correction of artifacts in MRI \cite{Jin2017,Mani2016b,Lee2016a,Mani2016a}.  As far as we know, this is the first work that use the idea of SLRMC for radial trajectory correction. The robustness of the reconstruction was tested using simulations and in-vivo data and compared to state-of-the-art methods. The proposed method can be extended to 3D and 4D radial trajectory corrections also. However, the high memory demands of the present method would require the adaption of memory-efficient structured low rank methods such as \cite{Ongie,Ongie2016b} for efficient implementation.

\par Due to their data-driven nature, the self-calibrating methods are more robust to temporal variations in trajectory errors compared to calibration-based correction methods. There are two potential ways in which these methods can be used for dynamic scan corrections. In the first approach, the trajectory correction can be performed on the first time point and the null space learned from this can be applied to the other time points of the dynamic scan. This approach may be prone to errors if the null space learned from the first time point may not be valid for a later time point; however it can be re-learned from a closer scan, similar to the idea of interleaved calibration scans for dynamics scanning \cite{Block}. A second approach is to perform independent trajectory correction for each time point at the expense of increased computation time.

\par Another advantage of the proposed scheme as a general phase compensation strategy is that this formulation allows for compensation of phase errors regardless of their source. Hence parametrization of the phase errors is not required in our approach.  Regardless, the proposed method is shown to work well for in-vivo data, where we do not know the characteristics of the phase errors present in the data beforehand. Our only assumption is that radial spokes with similar angles exhibit similar errors. While the partial Fourier data provided good correction with a 4-segment split, the conventional radial and golden angle full Fourier data provided good correction with a 6-segment and an 8-segment split respectively. Based on simulation studies, we observe that it is possible to get good reconstructions with more than one scheme of segmentation. It is intuitive to see that if the number of segments chosen are a multiple of the number of phases simulated, the proposed phase compensation constraint should still hold. Thus, as long as the trajectory shifts are within the band-limits of the implicit filter (parametrized by the size of the window chosen to form the Hankel matrix), the phase compensation can be effectively achieved. For small trajectory shifts this should be valid and we might be able to find band-limited annihilating filters for several choices of segments.

\par {Other structured low-rank based self-calibrating ghost correction methods have been proposed for EPI trajectories \cite{Lee2016a,Lobos2017}. In addition to the low-rank constraint, \cite{Lee2016a} makes use of a sparsity constraint to recover ghost-corrected EPI data on a channel-by-channel basis.  In contrast, our formulation imposes a parallel-imaging based data consistency to recover images. While this makes the proposed method applicable to only multi-coil acquisitions, the cardiac data in Figure 4 use only 3 channels in the reconstruction. Hence the number of coils  is not critical for good performance of the algorithm. At the same time, this makes it possible to work with accelerated acquisitions also efficiently. \cite{Lobos2017} has shown the utility of non-convex penalties for improved performance for EPI correction.}

\par In conclusion, we proposed a simple and generalizable method that formulates trajectory corrected reconstruction as a phase compensation problem. The method does not require explicit calibration of trajectory shifts or phase calibration and works for a variety of trajectories and accelerated acquisitions.

\clearpage

\newpage
{\small
\bibliographystyle{unsrtnat} 
\bibliography{ref_dti_cs_mrm}}

\newpage
 \begin{figure}[b!]
\begin{minipage}[l]{0.4\textwidth}
\caption{(a) Illustrative example showing an arbitrary k-space trajectory with 4 discrete k-space shifts. (b) The k-space data are separated into 4 matrices, $\widehat{I_1}[k]$ through $\widehat{I_4}[k]$. Hollow and solid ovals are used to represent the missing and known k-space data respectively. If all the k-space samples in each of the matrices were known, then the images corresponding to these matrices are phase-modulated versions of $\rho(\bf r)$. The real part of the images ${\cal{R}}\{I_i(\bf r)\}$ are shown to depict the phase modulations.  }
\subfigure[]{  \includegraphics[trim = 0mm 0mm 0mm 0mm, clip, width=1\textwidth]{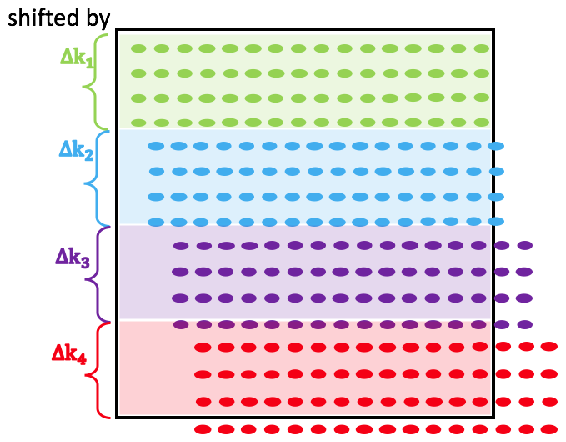}}
\end{minipage}
\begin{minipage}[l]{0.6\textwidth}
\subfigure[]{ \includegraphics[trim = 0mm 4mm 0mm 0mm, clip, width=.98\textwidth]{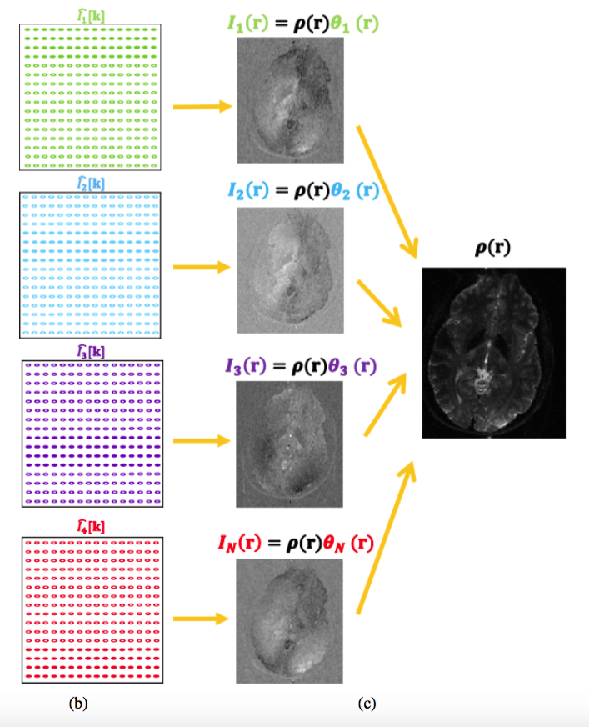}}
\end{minipage}
\label{fig:figS1}
\end{figure}
\clearpage

\newpage
 \begin{figure}
\includegraphics[trim = 0mm 0.5mm 0mm 0mm, clip, width=1\textwidth]{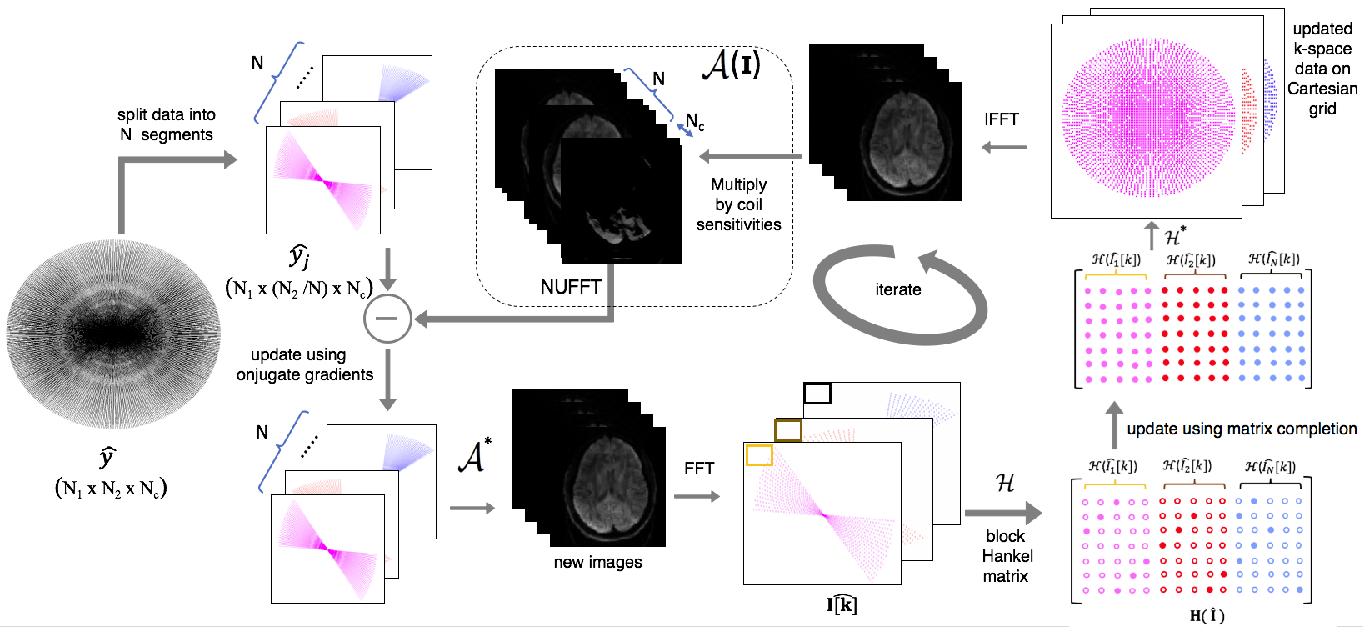}
\caption{The different steps involved in the proposed radial trajectory correction. The measured data is represented as $\mathbf {\hat y}$, which is split into N segments $\mathbf {\hat y_j}$. The operator $\mathcal A(\mathbf I)$ takes the images on the Cartesian grid, multiply those with the coil sensitivities and performs NUFFT to generate the k-space data corresponding to the radial k-space points; $\mathcal A^H$ performs the adjoint of $\mathcal A(\mathbf I)$. The operator $\mathcal H(\hat {\mathbf I})$ computes the Hankel matrix of the k-space data on the Cartesian grid and $\mathcal H^*(\hat {\mathbf I})$ performs the inverse mapping. }
\label{fig:figS1}
\end{figure}
\clearpage

\newpage
\begin{figure}[t!]
\includegraphics[trim = 4mm 3mm 3mm 2mm, clip, width=1\textwidth]{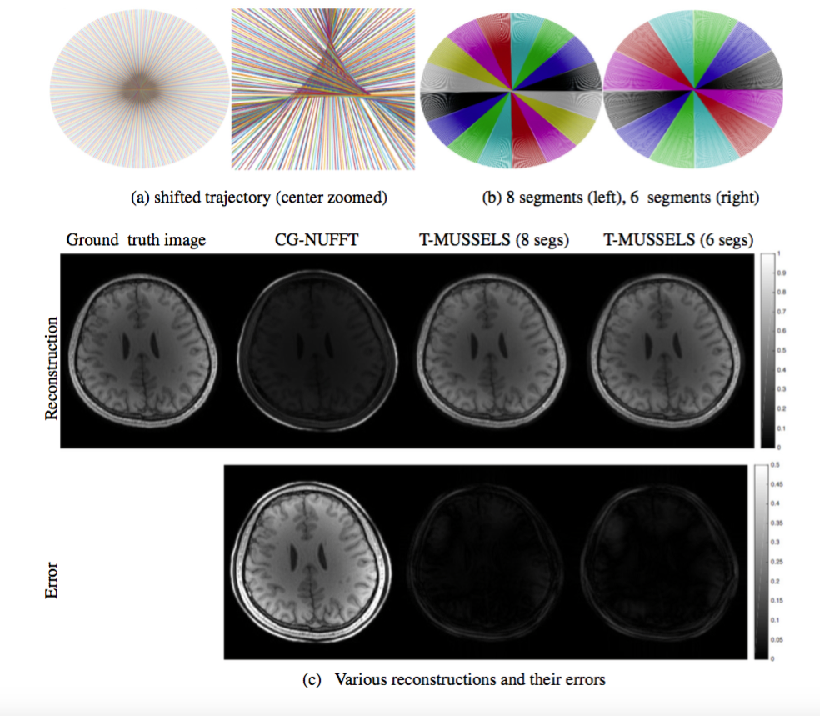}
\caption{Simulated experiments on radial trajectory. The shifted radial trajectory is shown in (a) and the nominal radial trajectory that was split into 8 and 6 segments for testing T-MUSSELS are shown in (b). The ground truth image is provided in (c). The CG-NUFFT reconstruction using nominal trajectory show very high error. The T-MUSSELS reconstruction can provide trajectory corrected reconstruction using 8 segments (nrmse=0.031) and 6 segments (nrmse=0.034).}
\label{fig:sim_epi_radial}
\end{figure}
\clearpage

\newpage
\begin{figure}[t!]
\begin{minipage}[l]{0.34\textwidth}
\caption{Trajectory error compensated reconstruction for in-vivo short-axis cardiac radial data from two cardiac frames.  The reconstructions time for CG-NUFFT, TrACR and T-MUSSELS are 2.54~sec, 356.76~sec and 166.14~sec respectively.}
\label{fig:cardiac_radial}
\end{minipage}
\begin{minipage}[l]{0.65\textwidth}
\includegraphics[trim = 3mm 1mm 1mm 1mm, clip, width=1\textwidth]{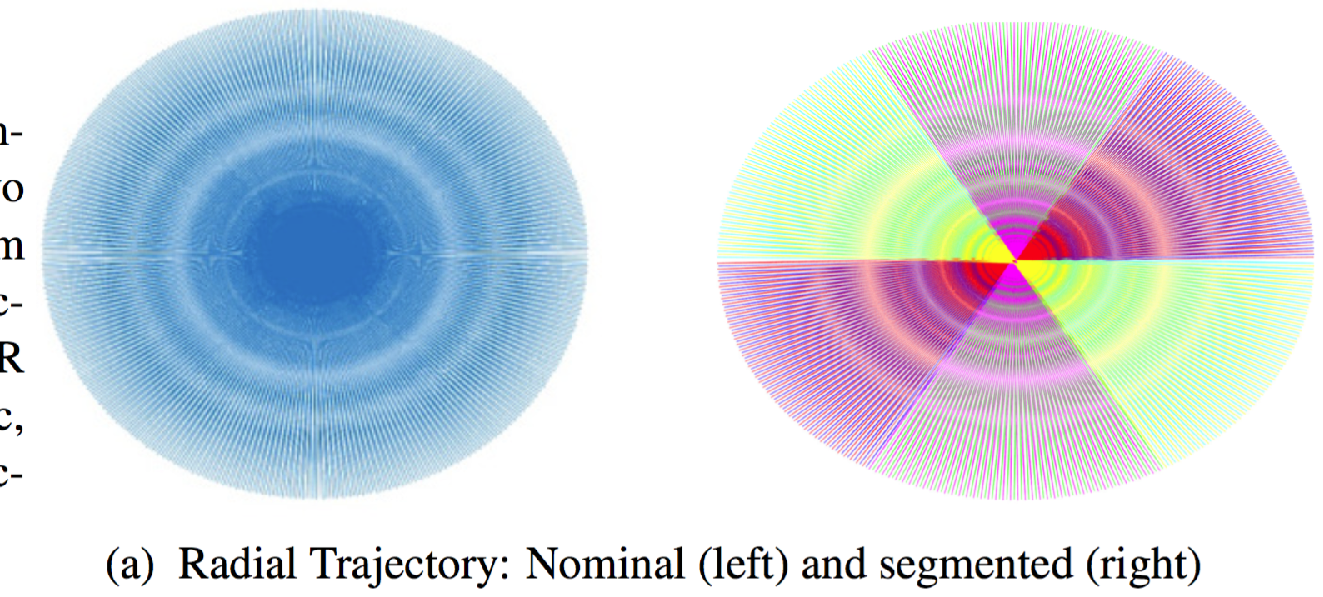}
\end{minipage}\\
\includegraphics[trim = 5mm 2.5mm 3mm 0mm, clip, width=1\textwidth]{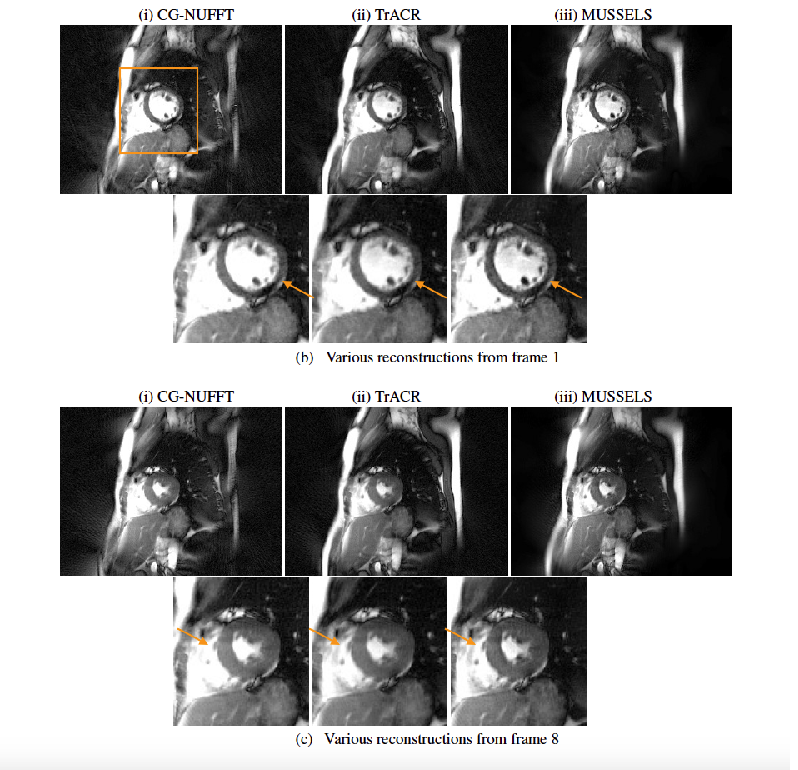}
\end{figure}
\clearpage

\newpage
\begin{figure}[t!]
\includegraphics[trim = 3mm 4mm 5mm 4mm, clip, width=1\textwidth]{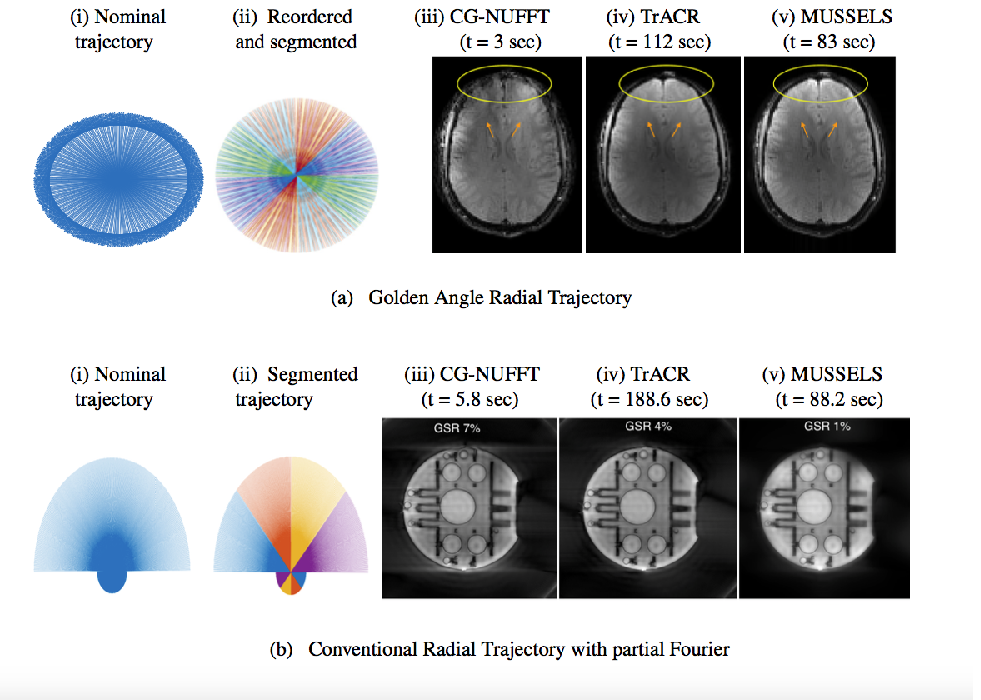}
\caption{Trajectory error compensated reconstruction for (a) golden angle radial data and (b) conventional radial with partial Fourier.  }
\label{fig:GA_radial}
\end{figure}
\clearpage

\end{document}